\documentclass[aps,pre,twocolumn,superscriptaddress,showpacs]{revtex4}
\usepackage{amsfonts,amssymb,amsmath,latexsym,epsfig} 
\begin{document}

[Phys. Rev. E {\bf 71}, 036215 (2005)]

\title{Effective dynamics in Hamiltonian systems with mixed phase space}

\author{Adilson E. Motter}
\email{motter@mpipks-dresden.mpg.de}
\affiliation{Max Planck Institute for the Physics of Complex Systems,
N\"othnitzer Strasse 38, 01187 Dresden, Germany}

\author{Alessandro P. S. de Moura}
\affiliation{Instituto de F\'{\i}sica, Universidade de S\~ao Paulo,
Caixa Postal 66318, 05315-970, S\~ao Paulo, SP, Brazil}

\author{Celso Grebogi}
\affiliation{Max Planck Institute for the Physics of Complex Systems,
N\"othnitzer Strasse 38, 01187 Dresden, Germany}
\affiliation{Instituto de F\'{\i}sica, Universidade de S\~ao Paulo,
Caixa Postal 66318, 05315-970, S\~ao Paulo, SP, Brazil}

\author{Holger Kantz}
\affiliation{Max Planck Institute for the Physics of Complex Systems,
N\"othnitzer Strasse 38, 01187 Dresden, Germany}

\date{\today}

\begin{abstract}

An adequate characterization of the dynamics of Hamiltonian systems at
physically relevant scales has been largely lacking.  Here we investigate this
fundamental problem and we show that the finite-scale Hamiltonian dynamics is
governed by effective dynamical invariants, which are significantly different
from the dynamical invariants that describe the asymptotic Hamiltonian dynamics.
The effective invariants depend both on the scale of resolution and the region
of the phase space under consideration, and they are naturally interpreted
within a framework in which the nonhyperbolic dynamics of the Hamiltonian system
is modeled as a chain of hyperbolic systems.

\end{abstract}

\pacs{05.45.-a, 05.45.Df}

\maketitle

\section{Introduction}
\label{1}

A comprehensive understanding of Hamiltonian dynamics is a long outstanding
problem in nonlinear and statistical physics, which has important applications
in various other areas of physics.  Typical Hamiltonian systems are
nonhyperbolic as they exhibit mixed phase space with coexisting regular and
chaotic regions.  Over the past years, a number of ground-breaking works
\cite{chirikov1,asymp,meiss1,greene1,pikovsky1,christiansen1,lau1,zasl,uptodate} have
increasingly elucidated the asymptotic behavior of such systems and it is now
well understood that, because of the stickiness due to Kolmogorov-Arnold-Moser
(KAM) tori, the chaotic dynamics of typical Hamiltonian systems is fundamentally
different from that of hyperbolic, fully chaotic systems.  Here ``asymptotic''
means in the limit of large time scales and small length scales.  But in
realistic situations, the time and length scales are limited.  In
the case of hyperbolic systems, this is not a constraint because the
(statistical) self-similarity of the underlying invariant sets guarantees the
fast convergence of the dynamical invariants (entropies, Lyapunov exponents,
fractal dimensions, escape rates, etc) and the asymptotic dynamics turns out to
be a very good approximation of the dynamics at finite scales.  In nonhyperbolic
systems, however, the self-similarity is usually lost because the invariant sets
are not statistically invariant under magnifications.  As a result, the
finite-scale behavior of a Hamiltonian system may be fundamentally different
from the asymptotic behavior considered previously, which is in turn hard to
come by either numerically \cite{uptodate,fluid} or experimentally \cite{nh_h}.

The aim of this paper is to study the dynamics of Hamiltonian systems at
finite, physically relevant scales. To the best of our knowledge, this problem
has not been considered before.  Herewith we focus on Hamiltonian chaotic
scattering, which is one of the most prevalent manifestations of chaos in open
systems, with examples ranging from fluid dynamics \cite{fluid,nh_h} to
solid-state physics \cite{sol_stat} to general relativity \cite{relat}.  We show
that the finite-scale dynamics of a Hamiltonian system is characterized by {\it
effective} dynamical invariants (e.g., effective fractal dimension), which:  (1)
may be significantly different from the corresponding invariants of the
asymptotic dynamics; (2) depend on the resolution but can be regarded
as constants over many decades in a given region of the phase space; and (3) may
change drastically from one region to another of the {\it same} dynamically
connected (ergodic) component.  These features are associated with the slow and
nonuniform convergence of the invariant measure due to the breakdown of
self-similarity in nonhyperbolic systems.  To illustrate the mechanism behind
the properties of the effective invariants, we introduce a simple deterministic
model which we build on the observation that a Hamiltonian system can be
represented as a chain of hyperbolic systems.

The paper is organized as follows.  We start, in Sec.  \ref{s2}, with the
analysis of the invariant measure and the outline of the transport structures
underlying its convergence.
Our chain model is introduced and analyzed in Sec.  \ref{3}.  The
effective fractal dimension is defined in Sec.  \ref{4} and its properties are
verified for a specific system in Sec.  \ref{5}.  Conclusions are presented
in the last section.

\section{Invariant Measure}
\label{s2}

\begin{figure}[pt]
\begin{center}
\epsfig{figure=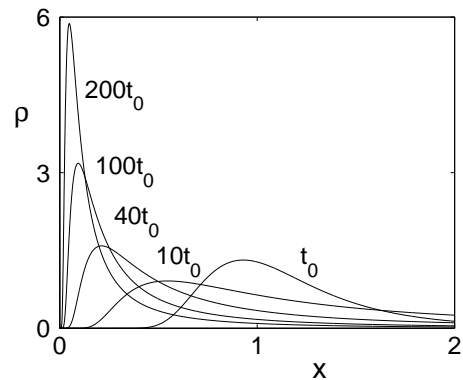,width=6.0cm}
\caption{Snapshots of the probability density $\rho$ as a
function of $x$, for $\rho(x,0)=\delta(x-x_0)$,
$x_0=1$, $x_1=2$, $\alpha=3$, and the outermost torus of
the KAM island at $x=0$ [APPENDIX].
The time $t$ is indicated in the figure.}
\label{fig1}
\end{center}
\end{figure}

For concreteness, consider a two-dimensional area preserving map with a major
KAM island surrounded by a chaotic region.  One such map captures all the main
properties of a wide class of Hamiltonian systems with mixed phase space.  When
the system is open (scattering), almost all particles initialized in the chaotic
region eventually escape to infinity.  We first study this case with a diffusive
model for the transversal motion close to the main KAM island, obtaining an
analytical expression for the probability density $\rho(x,t)$
of particles remaining in the scattering region at time $t$ and distance $x$
from the island [see APPENDIX].  We find that, in the case of chaotic scattering, a singularity
develops and the invariant measure, given by
$\lim_{t\rightarrow\infty}\rho(x,t)$, accumulates on the outermost KAM torus of
the KAM island [APPENDIX].  Physically, this corresponds to the tendency
of nonescaping particles to concentrate around the regular regions.
Dynamically, the stickiness due to KAM tori underlies two major features of
Hamiltonian chaotic scattering, namely the algebraic decay of the survival
probability of particles in the scattering region
\cite{asymp,meiss1,greene1,pikovsky1,christiansen1} and the integer
dimension of the chaotic saddle \cite{lau1}, and distinguishes this phenomenon
from the hyperbolic chaotic scattering characterized by exponential decay and
noninteger fractal dimension.  However, the convergence of the measure is rather
slow and highly nonuniform, as shown in Fig.~\ref{fig1} for typical parameters,
which is in sharp contrast with the fast, uniform convergence observed in
hyperbolic systems.  Our main results are ultimately related to this slow and
nonuniform convergence of the invariant measure.

Previous works on transport in Hamiltonian systems have used stochastic models,
where invariant structures around KAM islands are smoothened out and the
dynamics is given entirely in terms of a diffusion equation
\cite{chirikov1,greene1} or a set of transition probabilities (Markov chains or
trees) \cite{meiss1,other_chains}.  The stochastic approach is suitable to
describe transport properties (as above), but cannot be used to predict the
behavior of dynamical invariants such as Lyapunov exponents and fractal
dimensions.  Here we adopt a deterministic approach where we use the Cantori
surrounding the KAM islands to split the nonhyperbolic dynamics of the
Hamiltonian system into a chain of hyperbolic dynamical systems.

Cantori are invariant structures that determine the transversal transport close
to the KAM islands  \cite{asymp,meiss1}.  There is a hierarchy of infinitely many
Cantori around each island. Let $C_1$ denote the area of the
scattering region outside the outermost Cantorus, $C_2$ denote the annular area
in between the first and second Cantorus, and so on.  As $j$ is increased, $C_j$
becomes thinner and approaches the corresponding island.  For simplicity, we
consider that there is a single island \cite{hier} and that, in each iteration,
a particle in $C_j$ may either move to the outer level $C_{j-1}$ or the inner
level $C_{j+1}$ or stay in the same level \cite{meiss1}.  Let $\Delta_j^{-}$ and
$\Delta_j^{+}$ denote the transition probabilities from level $j$ to $j-1$ and
$j+1$, respectively.  A particle in $C_1$ may also leave the scattering region,
and in this case we consider that the particle has escaped.  The escaping region
is denoted by $C_0$.  The chaotic saddle is expected to have points in $C_j$ for
all $j\geq 1$.  It is natural to assume that the transition probabilities
$\Delta_j^{-}$ and $\Delta_j^{+}$ are constant in time.  This means that each
individual level can be regarded as a hyperbolic scattering system, with its
characteristic exponential decay and noninteger chaotic saddle dimension.
Therefore, a nonhyperbolic scattering is in many respects similar to a sequence
of hyperbolic scatterings.

\section{Chain Model}
\label{3}

\begin{figure}[bt]
\begin{center}
\epsfig{figure=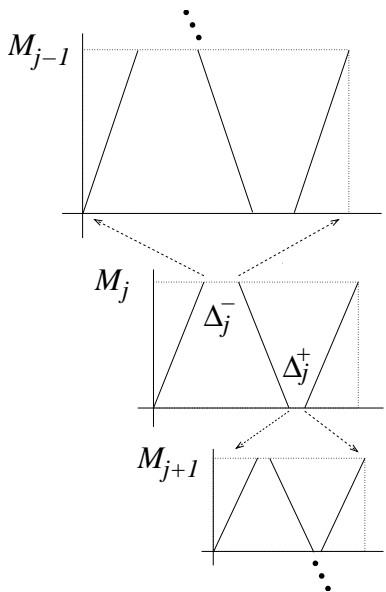,width=5.0cm}
\caption{Semi-infinite chain of hyperbolic maps $M_j$, $j=1,2,\ldots$}
\label{fig2}
\end{center}
\end{figure}

\begin{figure}[pt]
\begin{center}
\epsfig{figure=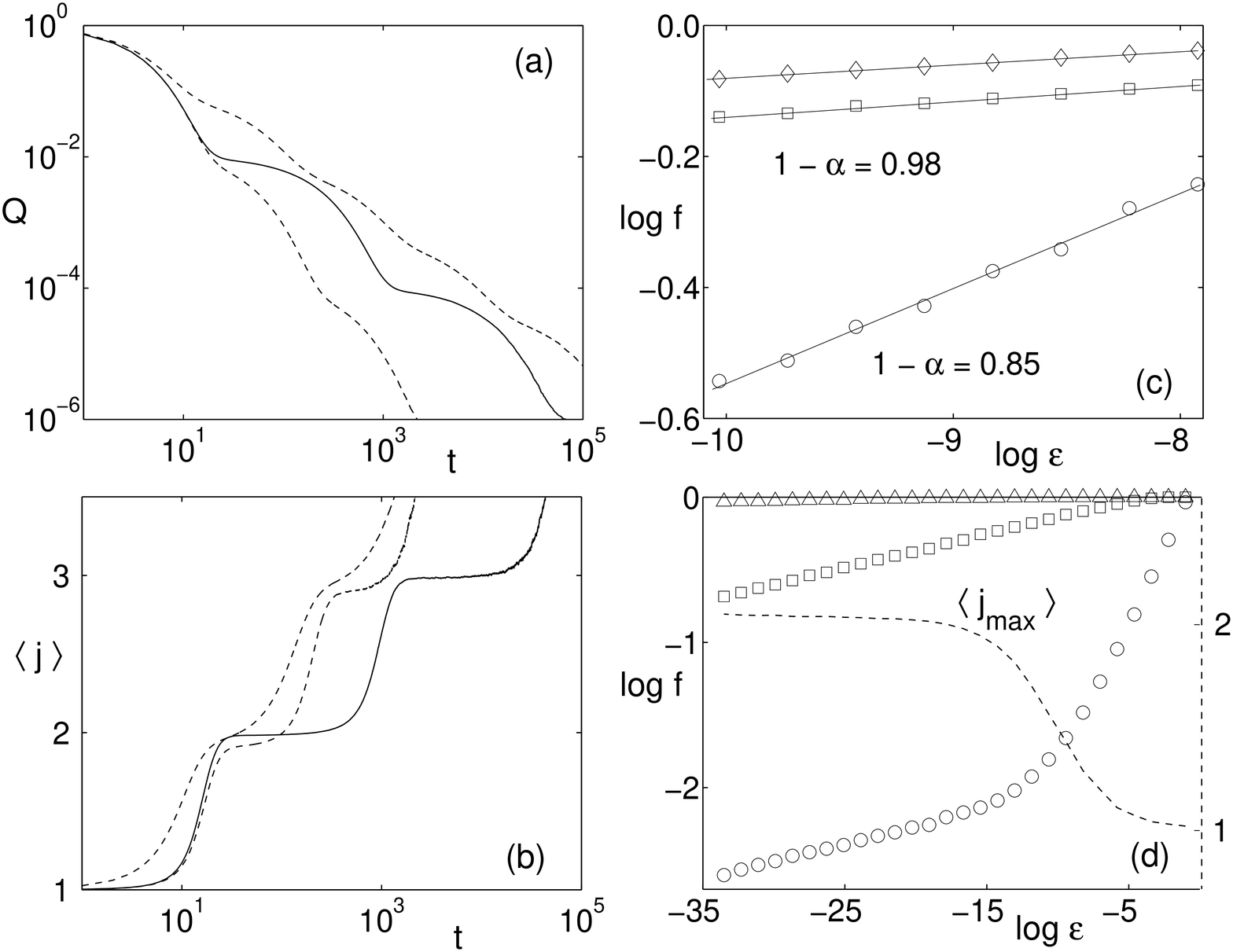,width=8.0cm}
\caption{Chain model for $\xi_1=4.1$.
         (a) Survival probability $Q$  and (b) average position $\langle j\rangle$ as a function of time
         for $\mu=0.01$ and $\nu=0.02$ (full line), $\mu=0.01$ and $\nu=0.1$ (dashed, bottom),
         and $\mu=0.08$ and $\nu=0.1$ (dashed, top).
         (c) Fraction $f(\varepsilon)$ of uncertain points as a
         function of the scale $\varepsilon$ for points taken from $L=I_1$ (circles), $L=I_2$ (squares),
         and the subinterval $L$ of $I_1$ first mapped into $I_2$ (diamonds),
         where $\mu=0.01$ and $\nu=0.1$.
         Circles in (c) are shifted vertically upward for clarity.
         (d) The same as in (c) for $\varepsilon \geq 10^{-35}$ and $L=I_1$ (circles), $L=I_2$ (squares),
         and $L=I_3$ (triangles). Dashed line (right-side axis): average maximum $j$  of 
         orbits started from $\varepsilon$-uncertain points, for $L=I_1$.
         }
\label{fig3}
\end{center}
\end{figure}

We now introduce a simple deterministic model that incorporates
the above elements and reproduces essential features of the Hamiltonian dynamics.
Our model is depicted in Fig.~\ref{fig2} and consists of a semi-infinite
chain of 1-dimensional ``$/\backslash/$-shaped'' maps,
defined as follows:
\begin{equation}
M_j(x) = \left \{
\begin{array}{lll}
  &\xi_j x,                       & 0\leq x<1/\xi_j\\
- &\xi_j (x - \Delta_j^{-}) +2,   & 1/\xi_j< x - \Delta_j^{-} <  2/\xi_j\\
  &\xi_j (x-1)+1,                 & -1/\xi_j < x -1 \leq 0 ,
\end{array}  \right. \nonumber
\end{equation}
where $\xi_j> 3$ and $0< \Delta_j^{-} < 1-3/\xi_j\;$
($j=1,2,\ldots$). If $x$ falls in the interval $ 1/\xi_j\leq x \leq
1/\xi_j + \Delta_j^{-}$, where $M_j$ is not defined, the ``particle''
is considered to have crossed a Cantorus to the ``outer level''
$j-1$. This interval is mapped uniformly to $[0,1]$, and the iteration
proceeds through $M_{j-1}$. Symbolically, this is indicated by
$j\rightarrow j-1$. Similarly, if $x$ falls into $ 1-1/\xi_j -
\Delta_j^{+}\leq x\leq 1-1/\xi_j$, where $\Delta_j^{+} = 1-3/\xi_j -
\Delta_j^{-}$, the particle goes to the ``inner level'', and
$j\rightarrow j+1$.
Particles
that reach $ 1/\xi_1\leq x
\leq 1/\xi_1 + \Delta_1^{-}$ are considered to have escaped. 
The domain of $M_j$ is denoted by $I_j$ and is analogous to $C_j$ in a Hamiltonian system,
where $\Delta_j^{-}$ and $\Delta_j^{+}$ represent the transition probabilities.
 The transition rate ratios
$\mu = \Delta_j^{+}/\Delta_j^{-}$ and $\nu = \Delta_{j+1}/\Delta_j$ are taken in
the interval $0 <\mu < \nu <1$ and are set to be independent of $j$, where
$\Delta_{j}= \Delta_j^{+} + \Delta_j^{-}$.
The parameter $\mu$ is a measure of the fraction of particles
in a level $j$ that will move to the inner level $j+1$ when leaving level $j$, while $\nu$
is a measure of how much longer it takes for the particles in the inner level to escape. 
The nondependence on $j$ corresponds to
the approximate scaling of the Cantori suggested by the renormalization theory \cite{meiss1}.
Despite the hyperbolicity of each map, the entire chain behaves as a nonhyperbolic system.
For a uniform initial distribution in $I_1$,
it is not difficult to show \cite{details} that
the number of particles remaining in the chain after a long time $t$
decays algebraically as $Q(t)\sim t^{-\ln \mu /\ln \nu}$,
and that the initial conditions of never escaping particles form a zero Lebesgue
measure fractal set
with box-counting  dimension 1.
However, the finite-scale behavior may deviate considerably from these
asymptotics, as shown in Fig.~\ref{fig3}.

In Fig.~\ref{fig3}(a) we show the survival probability $Q$ as a function of
time.  For small $\mu$ and $\nu$, the curve is composed of a discrete sequence
of exponentials with scaling exponents $\ln (1-\Delta_j^-)$,
which decrease  (in absolute value) as we go forward in the sequence. 
The length of each exponential segment is of the
order of $\mu$ in the decay of $Q$ and $-\ln\nu$ in the variation of $\ln t$.
This striking behavior is related to the time evolution of the density of
particles inside the chain.  This is shown in Fig.~\ref{fig3}(b), where we plot
the average position $\langle j\rangle$ of an ensemble of particles initialized
in $I_1$ (i.e., $j=1$).  The transitions between successive exponentials in the
decay of $Q$ [Fig.~\ref{fig3}(a)] match the transitions from a level $j$ to the
next in the average position of the remaining particles [Fig.~\ref{fig3}(b)].
In a Hamiltonian system, the increase of $\langle j\rangle$ in time is related
to the development of the singular invariant measure anticipated in our
diffusion analysis [see Fig.~\ref{fig1}].  The piecewise exponential behavior
of $Q$ is smoothened out for large $\mu$ and $\nu>\mu$ [Figs.~\ref{fig3}(a) and
\ref{fig3}(b)].  

In Fig.~\ref{fig3}(c) we show the fractal dimension of the set
of initial conditions of never escaping particles as computed from the
uncertainty algorithm \cite{uncert}, which consists in measuring the scaling
of the fraction  $f(\varepsilon)$ of {\it $\varepsilon$-uncertain} points (initial points
whose escaping time is different from the escaping time
of points taken $\varepsilon$ apart).
The scaling is statistically well defined
over decades and the exponent $\alpha=\Delta \ln f(\varepsilon)/\Delta \ln
\varepsilon$ can be computed accurately.
However, the resulting dimension $1-\alpha$ is not only significantly smaller
than 1 but also depends critically
on the region $L$ of the phase space where it is computed.  The convergence of
the dimension is indeed so slow that it can only be noticed when observed over
very many decades of resolution, as shown in Fig.~\ref{fig3}(d) where data of
Fig.~\ref{fig3}(c) is plotted over 35 decades!  Initially smaller, the dimension
measured for $L=I_1$ approaches the dimension measured for $L=I_2$ as the scale
$\varepsilon$ is reduced beyond $10^{-15}$ (i.e. the corresponding curves in
Fig.~\ref{fig3}(d) become parallel).  As shown in Fig.~\ref{fig3}(d), this
behavior is related to a transition in the average innermost level $\langle
j_{max} \rangle$ reached by the particles launched from $\varepsilon$-uncertain
points.  As $\varepsilon$ is further reduced, new transitions are expected.  The
dimension measured in between transitions is mainly determined by the dimension
$D=\ln 3/\ln \xi_k$, $k=\langle j_{max} \rangle$, of the corresponding element
of the chain. For given $j$ and $\varepsilon$,
 the measured dimension is larger when
 $L$ is taken in a denser part of the invariant set, 
 such as in the subinterval of $I_1$ first mapped into $I_2$ [Fig.~\ref{fig3}(c); diamonds],
 because $\langle j_{max} \rangle$ is larger in these regions. 
In some regions, however, the measured dimension is quite different
from the asymptotic value even at scales as small as $\varepsilon =10^{-30}$.
This slow convergence of the dimension is due to the slow increase of $\langle j_{max} \rangle$,
which in a Hamiltonian system is related to the slow convergence of the invariant
measure [Fig.~\ref{fig1}].
The convergence is even slower for smaller $\mu$ and larger $\nu$.
Incidentally, the experimental measurements of the fractal dimension are usually
based on scalings over less than two decades \cite{avnir1}.  Therefore, at
realistic scales the dynamics is clearly not governed by the asymptotic
dynamical invariants.

\section{Effective Dynamical Invariants}
\label{4}

Our results on the chain model motivate us to introduce the concept of 
effective dynamical invariants.  As a specific example, we consider the
{\it effective} fractal dimension, which, for the intersection of a fractal set $S$
with a $n$-dimensional region $L$, we define as
\begin{equation}
\left. D_{eff}(L;\varepsilon)= n-\frac{d \ln f(\varepsilon')}{d\ln \varepsilon'}\right|_{\varepsilon'=\varepsilon},
\label{2}
\end{equation}
where $f(\varepsilon')= N(\varepsilon')/N_0(\varepsilon')$, and
$N(\varepsilon')$ and $N_0(\varepsilon')$ are the number of cubes of edge length
$\varepsilon'$ needed to cover $S\cap L$ and $L$, respectively \cite{prev_work}.
We take $L$ to be a generic segment of line [i.e., $n=1$ in Eq. (\ref{2})] intersected
by $S$ on a fractal set.
In the limit $\varepsilon\rightarrow 0$, we recover the usual box-counting
dimension $D=1-\lim_{\varepsilon\rightarrow 0}\Delta \ln f(\varepsilon)/\Delta
\ln \varepsilon$ of the fractal set $S\cap L$, which is known to be 1 for all
our choices of $L$.  However, for any practical purpose, the parameter
$\varepsilon$ is limited and cannot be made arbitrarily small (e.g., it cannot
be smaller than the size of the particles, the resolution of the experiment, and
the length scales neglected in modeling the system).  At scale $\varepsilon$ the
system behaves as if the fractal dimension were $D_{eff}(L;\varepsilon)$
(therefore ``effective'' dimension).  In particular, the final state sensitivity
of particles launched from $L$, with the initial conditions known within
accuracy $\varepsilon^*$, is determined by $D_{eff}(L;\varepsilon^*)$ rather
than $D$:  as $\varepsilon$ is variated around $\varepsilon^*$, the fraction of
particles whose final state is uncertain scales as
$\varepsilon^{1-D_{eff}(L;\varepsilon^*)}$, which is different from the
prediction $\varepsilon^{1-D}$.  This is important in this context because, as
shown in Fig.~\ref{fig3} (where the effective dimension is given by $1-\alpha$), 
the value of 
$D_{eff}(L;\varepsilon)$
may be significantly different from the
asymptotic value $D=1$ even for unrealistically small $\varepsilon$ and may also
depend on the region of the phase space.  Similar considerations apply to many
other invariants as well.

We now return to the Hamiltonian case.  Consider a scattering process in which
particles are launched from a line $L$ transversal to the stable manifold $W_s$
of the chaotic saddle.  Based on the construction suggested by the chain model,
it is not difficult to see that $W_s\cap L$ exhibits a hierarchical structure
which is not self-similar and is composed of infinitely many nested Cantor sets,
each of which is associated with the dynamics inside one of the regions $C_j$.
As a consequence, the effective dimension $D_{eff}(L;\varepsilon)$ in
Hamiltonian systems is expected to behave similarly to the effective dimension
in the chain model [Figs.~\ref{fig3}(c) and \ref{fig3}(d)].  In particular,
$D_{eff}(L;\varepsilon)$ is expected to display a strong dependence on $L$ and a
weak dependence on $\varepsilon$.

\section{Numerical Verification}
\label{5}

\begin{figure}[pt]
\begin{center}
\epsfig{figure=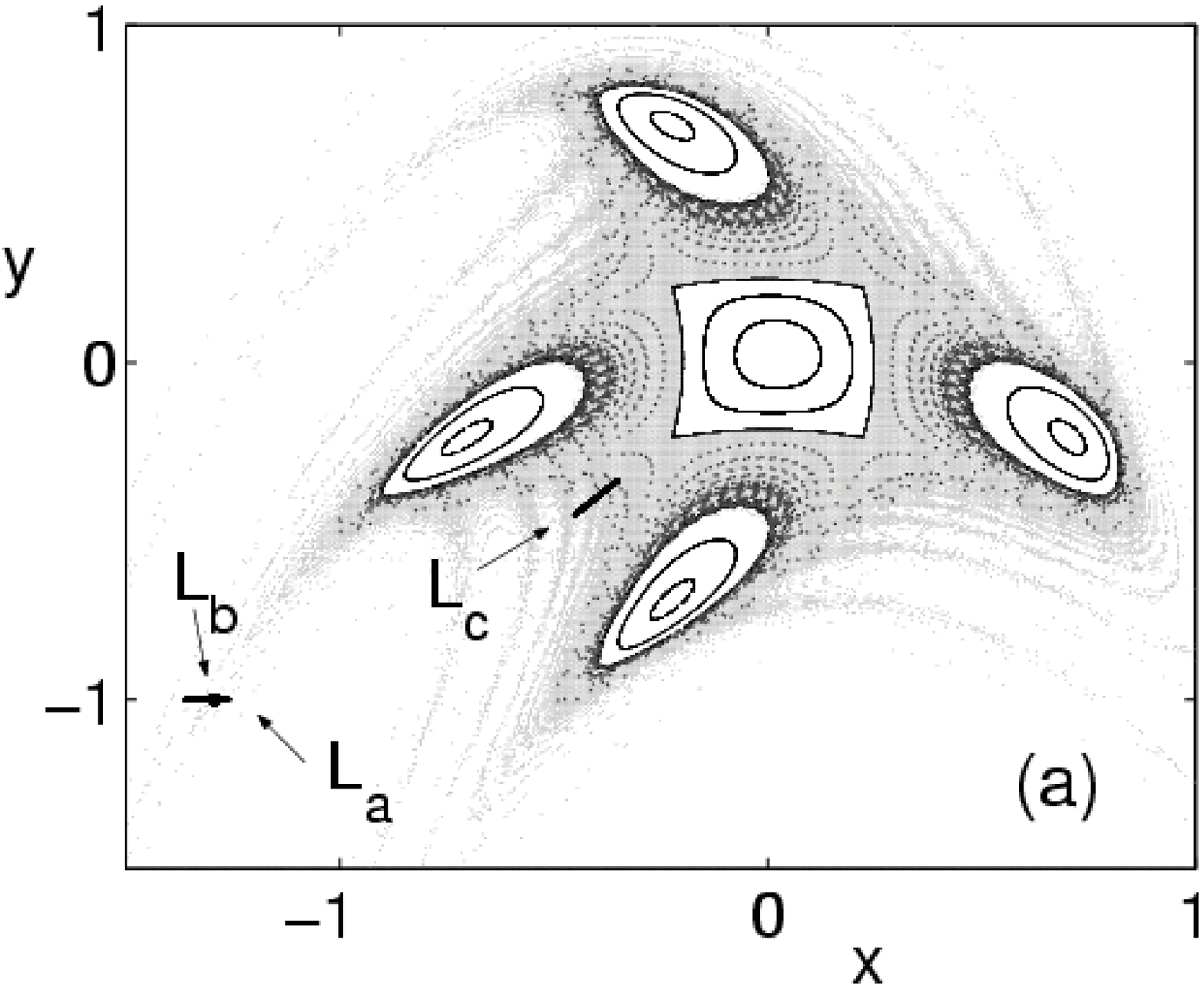,width=5.0cm}
\epsfig{figure=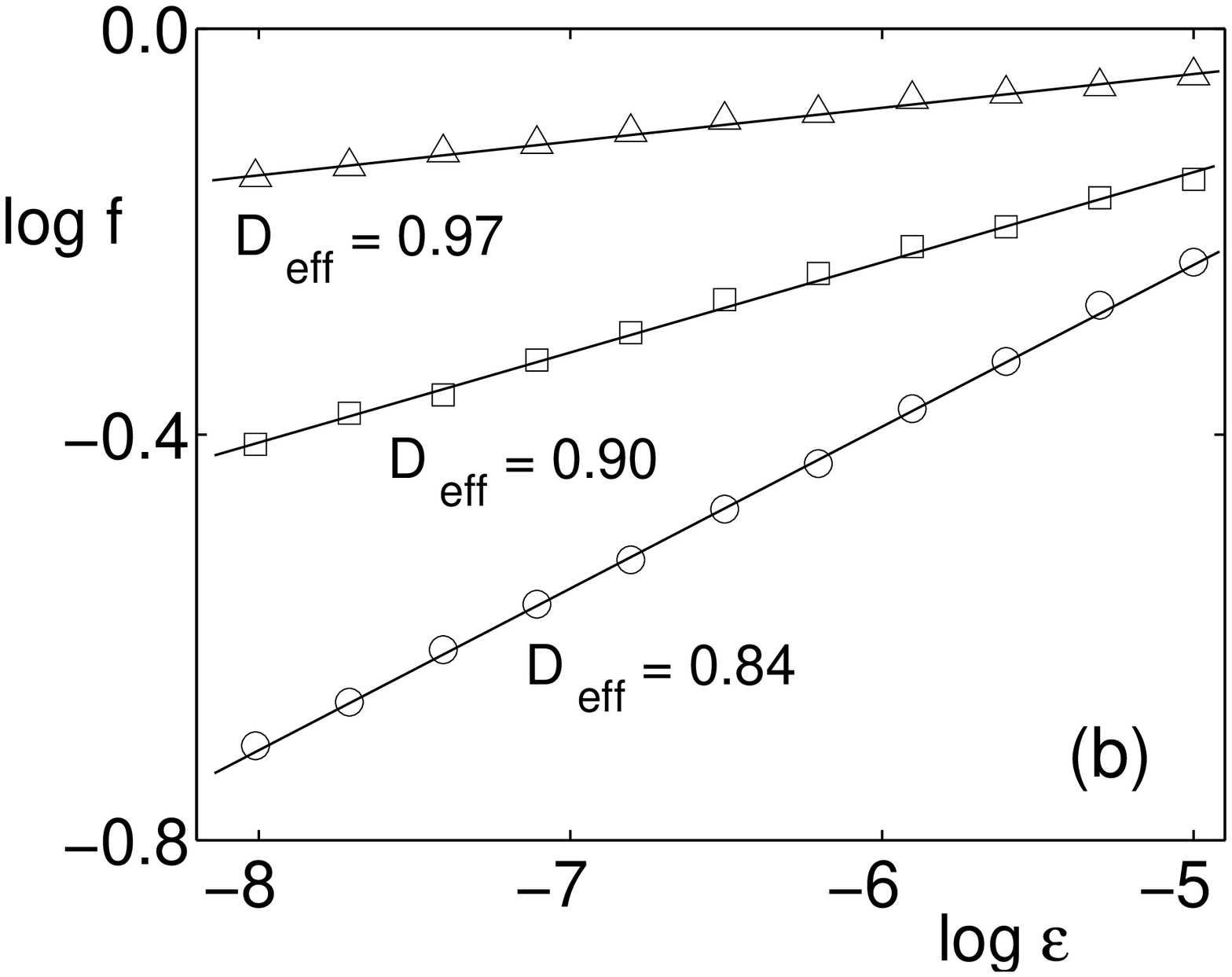,width=5.0cm}
\caption{(a) KAM islands (blank), stable manifold (gray),
             and the lines of initial conditions ($L_b$ is a subinterval of $L_a$). 
         (b) Effective dimension for $L=L_a$ (circles), $L=L_b$ (squares),
         and $L=L_c$ (triangles). The data in (b) are shifted vertically for clarity.}
\label{fig4}
\end{center}
\end{figure}

We test our predictions on the area preserving H\'enon map:  $f(x,y)=
(\lambda -y -x^2,x)$, where $\lambda$ is the bifurcation parameter.  In this
system, typical points outside KAM islands are eventually mapped to infinity.
Because of the symmetry $f^{-1}=g\circ f\circ g$, where $g(x,y)=(y,x)$,
the stable and unstable manifolds of the chaotic saddle are obtained from each other
by exchanging $x$ and $y$.
For $\lambda=0.05$, the system displays a period-one and a period-four major island, as shown in
Fig.~\ref{fig4}(a).  In the same figure we also show the complex invariant structure
around the islands, the stable manifold of the chaotic saddle, and three different
choices for the line of starting points:  a large interval away from the islands
($L_a$), a small subinterval of this interval where the stable manifold 
appears to be denser ($L_b$), and an interval closer to the islands ($L_c$).
The corresponding effective dimensions are computed
for a wide interval of $\varepsilon$.  The results are shown in Fig.~\ref{fig4}(b):
$D_{eff}(L_a;\varepsilon)= 0.84$,
$D_{eff}(L_b;\varepsilon)= 0.90$, and
$D_{eff}(L_c;\varepsilon)= 0.97$ for $10^{-8}<\varepsilon< 10^{-5}$.
These results agree with
our predictions that the effective fractal dimension has the following
properties:  $D_{eff}$ may be significantly different from the asymptotic
value $1$ of the fractal dimension; $D_{eff}$ depends on the
resolution $\varepsilon$ but is nearly constant over decades; $D_{eff}$ depends
on the region of the phase space under consideration and, in particular,
is larger in regions closer to the islands and in regions where the stable
manifold is denser.
Similar results are expected for any typical Hamiltonian system with mixed phase space.

\section{Conclusions}


We have shown that the finite-scale dynamics of Hamiltonian systems,
relevant for realistic situations, is governed by effective dynamical
invariants.  The effective invariants are not only different from the asymptotic
invariants but also from the usual hyperbolic invariants because they strongly
depend on the region of the phase space.  Our results are generic and expected
to meet many practical applications.  In particular, our results are expected to
be relevant for fluid flows, where the advection dynamics of tracer particles is
often Hamiltonian \cite{fluid}.  In this context, a slow nonuniform convergence
of effective invariants is expected not only for time-periodic flows, capable of
holding KAM tori, but also for a wide class of time-irregular incompressible
flows with nonslip obstacles or aperiodically moving vortices.

\acknowledgements

This work was supported by MPIPKS, FAPESP, and CNPq.
A. E. M. thanks Rainer Klages for illuminating discussions.

\appendix*
\section{}
\label{appendix}

The diffusion model is:  $\partial_t P(x,t) = \partial_x [x^{\alpha}\partial_x
P(x,t)]$, where $P$ is the probability density of all particles, $x\ge 0$, and
$\alpha>2$ \cite{chirikov1}.  The outermost torus of the KAM island is at $x=0$,
where the diffusion rate (proportional to $x^{\alpha}$) vanishes.  In a chaotic
scattering process the initial distribution of particles is localized apart from
the confining islands.  We take $P(x,0)=\delta (x-x_0)$, $x_0>0$, and consider a
particle to escape when it reaches $x\geq x_1$.  Under the approximation that
for large $x_1$ the return of particles can be neglected, we disregard the
boundary condition $P(x_1,t)=0$ and we take the solution to be the corresponding
Green function:  $P(x,t)=(\alpha-2) (xx_0)^{-1/2} yy_0 \exp(-y^2-y_0^2)
I_{\beta}(2yy_0)$, where $y=(\alpha-2)^{-1}t^{-1/2}x^{-(\alpha-2)/2}$, $y_0$ is
$y$ at $x=x_0$, $\beta=(\alpha-1)/(\alpha-2)$, and $I_{\beta}$ is the modified
Bessel function, which scales as $I_{\beta}\sim (2yy_0)^{\beta}$ for small
$2yy_0$ \cite{greene1}.  For any fixed $x>0$, we can show that the distribution
for large $t$ decreases as $P(x,t)\sim t^{-\beta-1}$, where
$\beta=(\alpha-1)/(\alpha-2)$.  On the other hand, as shown in Ref.~\cite{pikovsky1},
the fraction of particles in the interval $x<x_1$ decays
algebraically as $Q(t)\equiv\int_0^{x_1} P(x,t)$d$x\sim t^{-\beta}$.  Combining
these two results, it follows that the normalized probability density
$\rho(x,t)\equiv P(x,t)/Q(t)$ decreases as $\rho(x,t) \sim t^{-1}$ at each fixed
$x\in (0,x_1)$ for large enough $t$ and diverges arbitrarily close to $x=0$.



\begin{references}
\bibitem{hier}
Generalization to a full hierarchy of KAM
islands is straightforward and does not change the essence of our results.
\bibitem{prev_work}
Different effective invariants have been introduced in Refs.~\cite{finite1}.
See also Refs.~\cite{finite2}.
\end{references}
\end{document}